\documentclass[]{raa}
\usepackage{graphicx,times}
\usepackage{natbib}

\begin{document}

   \title{Hard X-ray emission cutoff in anomalous X-ray pulsar 4U 0142+61 detected by INTEGRAL}

   \volnopage{Vol.0 (201x) No.0, 000--000}      
   \setcounter{page}{1}          

   \author{Wei Wang$^1$, Hao Tong$^{2}$, and Yan Jun Guo$^{3}$}

   \institute{$^1$ National Astronomical Observatories, Chinese Academy of Sciences,
             Beijing 100012, China; {\it wangwei@bao.ac.cn}\\
             $^2$ Xinjiang Astronomical Observatory, Chinese Academy of Sciences, Urumqi, China\\
             $^3$ School of Physics, Peking University, Beijing, China}

   \date{Received~~2013 month day; accepted~~2013~~month day}

\abstract{The anomalous X-ray pulsar 4U 0142+61 was studied by the INTEGRAL observations. The hard X-ray spectrum of 18 -- 500 keV for 4U 0142+61 was derived using near 9 years of INTEGRAL/IBIS data. We obtained the average hard X-ray spectrum of 4U 0142+61 with all available data. The spectrum of 4U 0142+61 can be fitted with a power-law with an exponential high energy cutoff. This average spectrum is well fitted with a power-law of $\Gamma\sim 0.51\pm 0.11$ plus a cutoff energy at $128.6\pm 17.2$ keV. The hard X-ray flux of the source from 20 -- 150 keV showed no significant variations (within 20$\%$) from 2003 -- 2011. The spectral profiles have some variability in nine years: photon index varied from 0.3 -- 1.5, and cutoff energies of 110 -- 250 keV.  The detection of the high energy cutoff around 130 keV shows some constraints on the radiation mechanisms of magnetars and possibly probes the differences between magnetar and accretion models for these special class of neutron stars. Future HXMT observations could provide stronger constraints on the hard X-ray spectral properties of this source and other magnetar candidates.
 }

\authorrunning{W. Wang et al.}            
   \titlerunning{Hard X-ray cutoff in 4U 0142+61}

\maketitle

\section{Introduction}

Anomalous X-ray pulsars (AXPs) and soft gamma-ray repeaters (SGRs) are a special class of neutron stars.
They generally have the following observed features (Mereghetti 2008): spin period in the
range of 2 --12 seconds; young characteristic age (10-100 kyr); some
AXPs are associated with supernova remnants; an inferred
dipole magnetic field\footnote{The low magnetic field magnetar should be treated
separately (Rea et al. 2010; Tong \& Xu 2012).} of $10^{14}-10^{15}$ G;
persistent X-ray luminosity (0.1-10 keV) in the range of $10^{34}-10^{36} \,{\rm erg\
s^{-1}}$ which is higher than their spin-down power; in the soft X-ray
band, they have very soft spectra, $kT \sim 0.5$ keV, or the
photon index $\Gamma\sim 2-4$; recurrent bursts (including giant flares);
the absence of massive companion stars. At present, growing evidence indicates that AXPs and SGRs may be
isolated neutron stars with the ultrastrong magnetic field (higher
than the quantum critical magnetic field $B_{\rm QED}
=m_e^2c^3/e\hbar = 4.4\times 10^{13}$ G), so called magnetars (Thompson \& Duncan 1996).

Non-thermal hard X-ray emissions above 15 keV in AXPs were discovered by
some missions like INTEGRAL, RXTE and SWIFT (Kuiper et al. 2004, 2006; Wang 2008).
The hard X-ray properties of AXPs are quite different from their spectral properties in soft X-ray bands below 10 keV.
In soft X-ray bands, the spectra of AXPs are generally attributed to a thermal component with a temperature of $kT\sim 0.35-0.6$ keV,
and a soft power-law component with $\Gamma\sim 2-4$ (Enoto et al. 2010a).
While in hard X-ray band above 20 keV, AXPs show a non-thermal emission component with a photon index of $\Gamma \sim 0.5-1.5$,
and the pulse fraction is larger than $50\%$ or even near $100\%$ (Kuiper et al. 2004, 2006).
The hard X-ray emission luminosity similar to those of soft X-ray bands is also much higher than the spin-down power of AXPs.
Thus the hard X-ray emission component should come from other energy power like magnetar activity.
The physical mechanism to produce this non-thermal hard-ray emission is still unknown. In addition, the new discovery of the hard X-ray spectrum above 10 keV
showed no cut-off up to 100 keV, implying that the luminosity
above 10 keV could possibly be the dominated component over the
softer band, which provides a new challenge to the magnetar model.

Then the discovery of the spectral cut-off in hard X-rays is
very important to constrain the total energy power released from
AXPs, furthermore provides the strong constraints on the magnetar
model. Kuiper et al. (2006) use the archived CGRO/COMPTEL data to constrain the spectrum of four AXPs in MeV bands, suggesting that the hard X-ray emission could not extend to MeV ranges (maybe show a cutoff below 1 MeV). Wang (2008) obtained the hard X-ray spectra of AXPs with the 2-year SWIFT data from 15 -- 200 keV, suggesting possible cutoff energies around 100 -- 150 keV for hard X-ray emissions of 1E 1841-045 and 4U 0142+61. Up to now, the spectral cutoff properties of AXPs in hard X-rays are not well understood yet.

In this work, we will study 4U 0142+61, the brightest one of all AXPs in X-ray bands with INTEGRAL/IBIS. The soft X-ray spectrum of 4U 0142+61 from 0.5 -- 10 keV by XMM-Newton is well described by a blackbody with temperature 0.41 keV
plus a power law with photon index 3.88, and $N_{\rm H}\sim
1.0\times 10^{22}{\rm cm^{-2}}$ (Rea et al. 2007). According to
the extinction of its optical counterpart, Hulleman et al. (2000)
estimated that the pulsar must be at a distance $>2.7$ kpc. Above 20 keV, 4U 0142+61 shows a non-thermal spectrum with a photon index of $\sim 0.9 - 1$ with RXTE and SWIFT data (Kuiper et al. 2006; Wang 2008). den Hartog et al. (2008) used all available observations on 4U 0142+61 by INTEGRAL, RXTE, XMM-Newton, and ASCA to analyze X-ray spectral properties, and found a mean photon index of $\sim 0.9$ from 20 -- 150 keV. Suzaku also observed 4U 0142+61 up to 70 keV (Enoto et al. 2011), and suggested that the hard X-ray component can be fitted by a power-law of $\Gamma\sim 0.9$, and a possible cutoff energy $> 150$ keV. INTEGRAL/IBIS is a high sensitivity hard X-ray imager from 18 -- 500 keV. Now we have available IBIS observations from 2003 to 2011.  Deeper exposure observations on the source 4U 0142+61 will help to constrain the spectral shape of a typical AXP in hard X-ray to soft gamma-ray bands.

We will first introduce the INTEGRAL observations on the source and data analysis in \S 2. In \S 3, the hard X-ray spectral properties of the AXP 4U 0142+61 are presented, and we will concentrate on the spectral cutoff feature. The spectral variations of 4U 0142+61 from 2003 -- 2011 from 18 -- 500 keV were shown in \S 4. Finally the physical implications of this spectral cutoff will be discussed in \S 5. A brief conclusion is given in \S 6.

\begin{figure}
\centering
\includegraphics[angle=0,width=12cm]{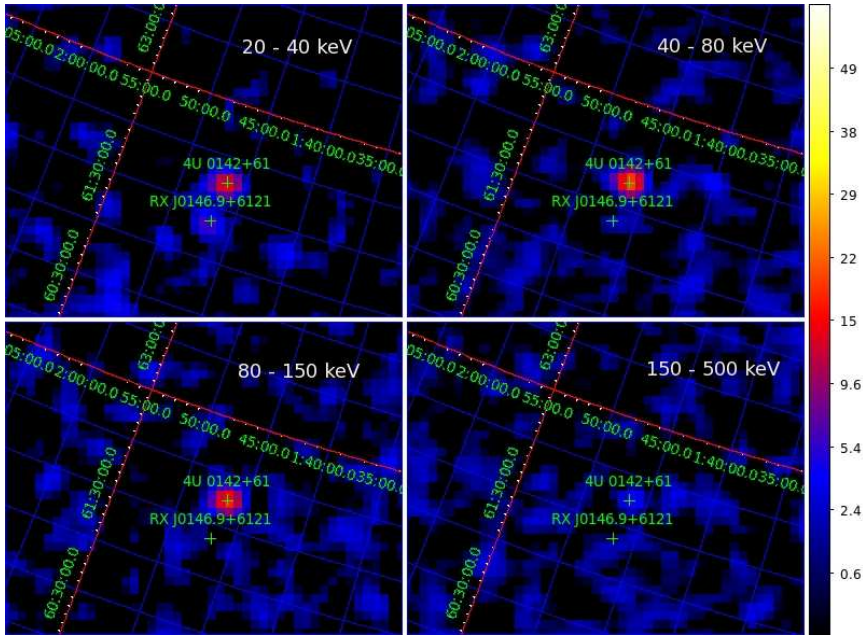}
\caption
{Significance mosaic maps around the AXP 4U 0142+61 in Equatorial J2000 coordinates as seen with INTEGRAL/IBIS in four energy bands: 20 -- 40 keV, detection significance level is $15.6\sigma$; 40 -- 80 keV, $20.9\sigma$; 80 -- 150 keV, $17.8\sigma$; 150 -- 500 keV, $<5\sigma$. }
\end{figure}

\section{INTEGRAL Observations}

4U 0142+61 was observed with frequently pointing observational surveys on Cassiopeia region from 2003 -- 2011 by the INTEGRAL satellite. We mainly use the observational data obtained by the INTEGRAL Soft Gamma-Ray Imager (IBIS-ISGRI, Lebrun et al. 2003) which has a 12' (FWHM)
angular resolution and arcmin source location accuracy in the energy band of 18 -- 500 keV.

We use the available archival data from the INTEGRAL Science Data Center (ISDC)
where 4U 0142+61 was within $\sim 12$ degrees of the pointing
direction of INTEGRAL/IBIS observations from 2003 -- 2011. The total corrected on-source time obtained in our analysis is about 3.8 Ms after excluding the bad data due to solar flares and the INTEGRAL orbital phase near the radiation belt of the Earth. The analysis was done with the standard INTEGRAL off-line scientific analysis (OSA, Goldwurn et al. 2003) software, ver. 10. Individual pointings in all collected IBIS data processed with OSA 10 were mosaicked to create the sky images for the source detection in given energy ranges.

In Fig. 1, sky images in four energy bands around 4U 0142+61 are displayed. In the energy range of 20 -- 40 keV, 4U 0142+61 is detected with a significance level of $15.6\sigma$, and a nearby source RX J0148.9+6121 is a high mass X-ray binary. In the range of 40 -- 80 keV, the source RX J0148.9+6121 cannot be detected by IBIS, while 4U 0142+61 is detected with a significance level of $20.9\sigma$. In the range of 80 -- 150 keV, 4U 0142+61 is still detected with a high significance level of $17.8\sigma$. But above 150 keV, the AXP cannot be detected by IBIS. Non-detection of 4U 0142+61 above 150 keV suggests the existence of possible high energy cutoff.

\section{Average Hard X-ray Spectrum of 4U 0142+61}

After the mosaicked images in the different energy ranges, we then can extract the average hard X-ray spectrum from 18 -- 500 keV by IBIS. The spectrum of 4U 0142+61 in the form of $E^2\times F$ obtained by IBIS is presented in Fig. 2. The spectral analysis software package used here is XSPEC 12.6.0q. In the spectral fittings, we have tried three simple spectral models to constrain the hard X-ray spectral properties of the AXP. In Table 1, we also show the spectral parameters fitted with the three models: a single power-law model, a thermal bremsstrahlung model and a power-law plus the exponential cutoff model. The bremsstrahlung model gives a temperature of $\sim 200$ keV, however cannot give a good fit on the spectrum (reduced $\chi^2$ is higher than 4). The single power-law model derives a photon index of $\sim 1$, but the fitting is not well satisfied. The spectral data points in the bands 18 -- 35 keV and above 150 keV can not be fitted well. The spectrum then is fitted with the cutoff power-law model, giving a photon index of $\sim 0.5$ and a spectral cutoff around $\sim 129\pm 17$ keV. The best fitted spectral model is also shown in Fig. 2.

\begin{figure}
\centering
\includegraphics[angle=-90,width=9.0cm]{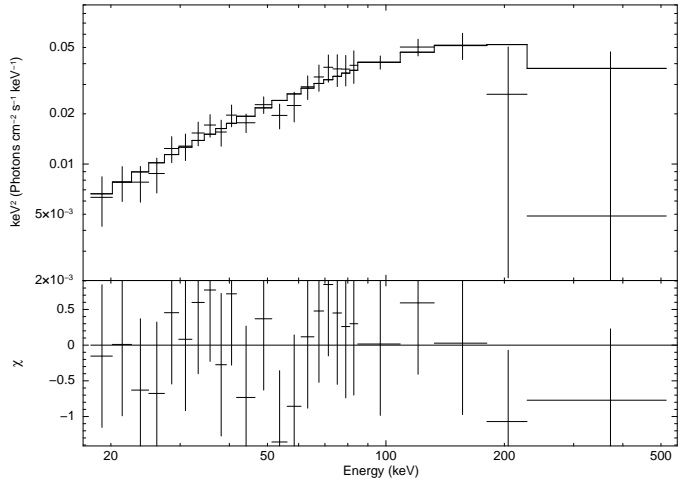}
\caption{The hard X-ray spectrum of 4U 0142+61 from 18 -- 500 keV obtained by INTEGRAL/IBIS. The spectrum is well fitted with a power-law plus an exponential cutoff model, with a photon index of $\sim 0.5$ and high energy cutoff energy of $\sim 130$ keV.  }

\end{figure}

\begin{table}

\caption{The hard X-ray spectral properties of the AXP 4U 0142+61 from 18 -- 500 keV. Three different spectral models are applied to fit the spectrum: a single power-law model, a thermal bremsstrahlung model and a power-law plus the exponential cutoff model. The flux is given in units of $10^{-10}$ erg cm$^{-2}$ s$^{-1}$ in the range of 18 -- 200 keV. The error bars are given in 1$\sigma$. }
\begin{center}
\begin{tabular}{l c c c l}
\hline \hline
Model &  $\Gamma/kT$ (keV)  &  Cutoff Energy (keV)  & Flux  &  reduced $\chi^2\ (d.o.f)$ \\
\hline
Power-law & $1.09\pm 0.07$ &  & $1.39\pm 0.11$ &  1.485 (24) \\
Bremsstrahlung & 179$\pm 58$ &  & $0.87\pm 0.12$ & 4.031 (24) \\
Cutoff power-law &  $0.51\pm 0.11$ & $128.6\pm 17.2$ & $1.27\pm 0.12$ & 0.614 (23) \\
\hline

\end{tabular}
\end{center}
\end{table}

Thus a simple spectral analysis with different models supports that the hard X-ray emissions from the brightest AXP 4U 0142+61 should have a spectral cutoff around 130 keV. The hard X-ray emission cannot extend to higher energy range like above 300 keV for AXPs with cutoff. From 18 -- 200 keV, the derived X-ray flux of 4U 0142+61 is about $(1.27\pm 0.12)\times 10^{-10}$ erg cm$^{-2}$ s$^{-1}$. This flux is lower than that given by den Hartog et al. (2008) which reported a hard X-ray flux of $(1.50\pm 0.08)\times 10^{-10}$ erg cm$^{-2}$ s$^{-1}$ from 20 -- 229 keV with a simple power-law fitting. In Table 1, the power-law model found a hard X-ray flux of $(1.39\pm 0.11)\times 10^{-10}$ erg cm$^{-2}$ s$^{-1}$ which is still consistent with the results by den Hartog et al. (2008). Assuming a distance of 3 kpc, the obtained total hard X-ray luminosity is $\sim 1.5\times 10^{35}$ erg s$^{-1}$.  This luminosity is still similar to
the total X-ray luminosity in softer bands (like 0.5 -- 10 keV, see Rea et al. 2007).

\section{Hard X-ray Spectral Variations of 4U 0142+61}

\begin{figure}
\centering
\includegraphics[angle=0,width=9.0cm]{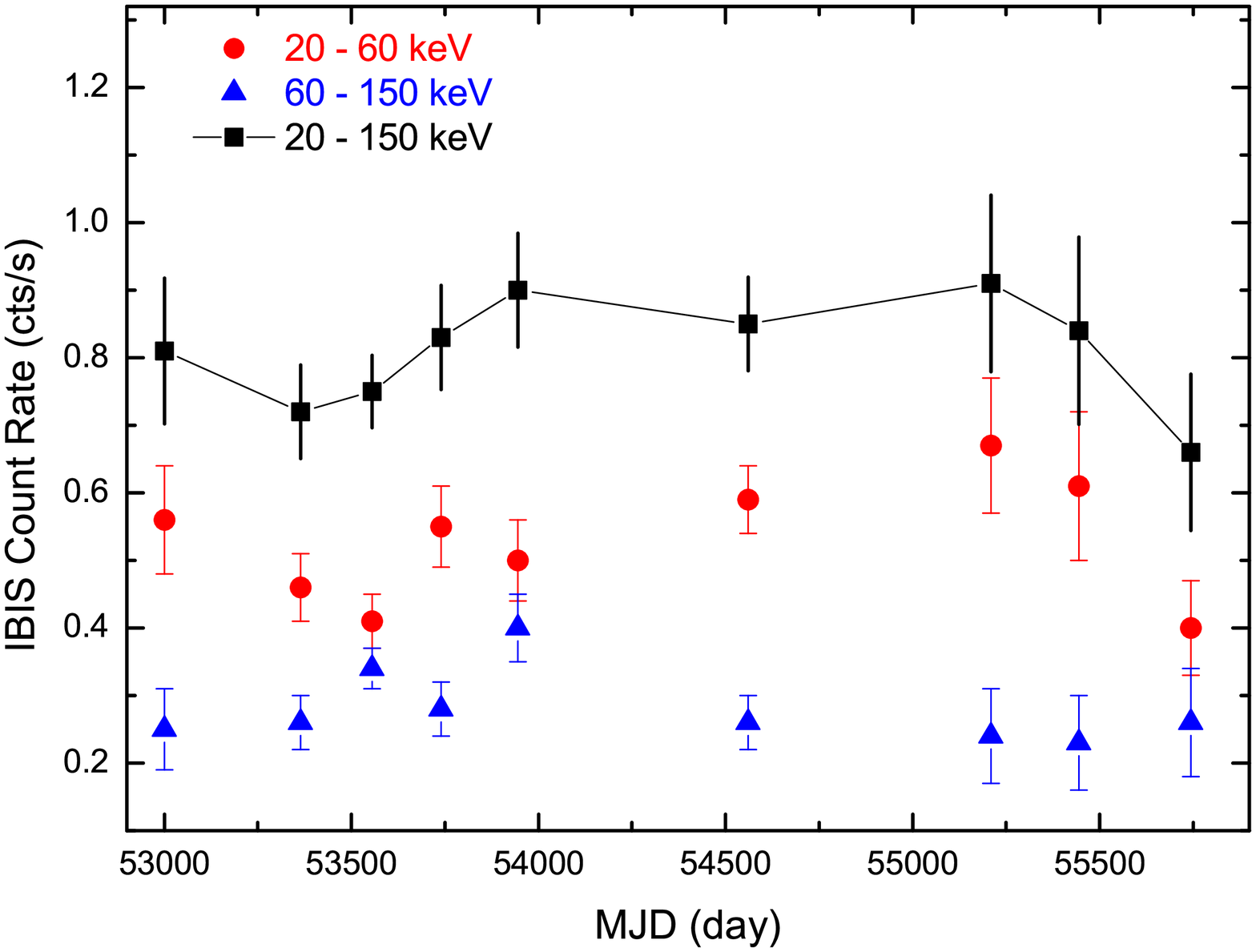}
\caption{The background subtracted IBIS count rate variations of 4U 0142+61 in energy bands: 20 -- 60 keV, 60 -- 150 keV and 20 -- 150 keV from 2003 -- 2011. The error bars are given with a confidence level of 1$\sigma$  }

\end{figure}

In \S 3, we derived the average spectrum of 4U 0142+61 with all available INTEGRAL observational data. This source or its spectral property could be variable in different time intervals (also see den Hartog et al. 2008). In Fig. 3, we present the IBIS count rate variations with years in two energy bands: 20 -- 60 keV and 60 -- 150 keV. From 2003 -- 2011, the total IBIS count rate of 4U 0142+61 from 20 -- 150 keV remains relatively stable around $\sim 0.7-0.9$ cts/s, the variations are less than $20\%$. Our result confirms that 4U 0142+61 is a stable and persistent hard X-ray emission source by previous INTEGRAL studies (Kuiper et al. 2006; den Hartog et al. 2008). But the count ratio of two energy bands (60 -- 150 keV to 20 -- 60 keV) changes with time, from 0.3 -- 0.9, suggesting that the spectral properties of 4U 0142+61 should vary with different time intervals.

\begin{figure}
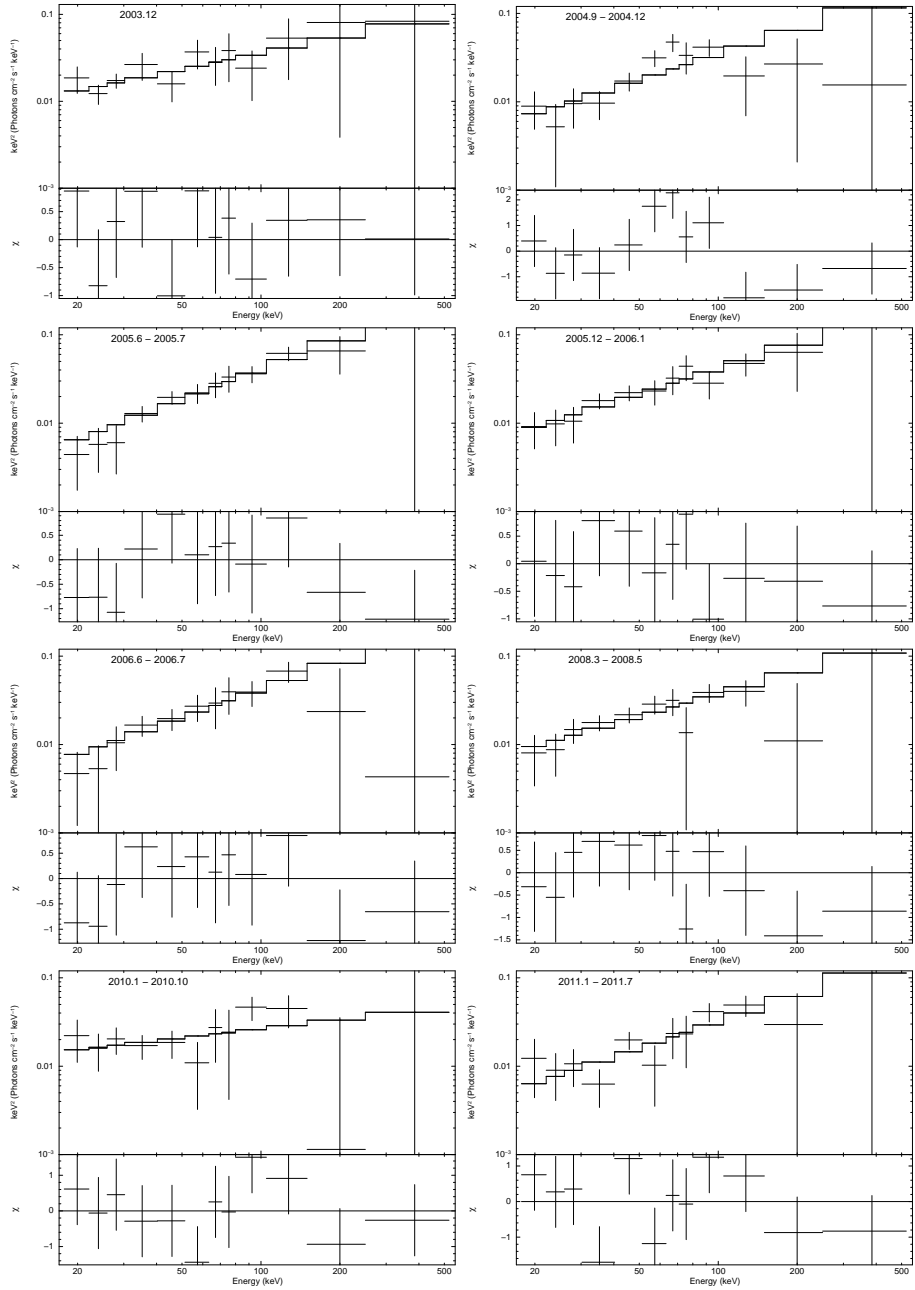

\centering
\includegraphics[angle=-90,width=6.0cm]{4u0142_2003o.eps}
\includegraphics[angle=-90,width=6.0cm]{4u0142_2004o.eps}
\includegraphics[angle=-90,width=6.0cm]{4u0142_2005o.eps}
\includegraphics[angle=-90,width=6.0cm]{4u0142_2006_1o.eps}
\includegraphics[angle=-90,width=6.0cm]{4u0142_2006_7o.eps}
\includegraphics[angle=-90,width=6.0cm]{4u0142_2008o.eps}
\includegraphics[angle=-90,width=6.0cm]{4u0142_2010o.eps}
\includegraphics[angle=-90,width=6.0cm]{4u0142_2011o.eps}
\caption{The hard X-ray spectra of 4U 0142+61 from 18 -- 500 keV obtained by INTEGRAL/IBIS in eight different time intervals. The spectra are all fitted with a simple power-law model whose parameters are presented in Table 2.  }

\end{figure}

Thus, we extracted the spectra of 4U 0142+61 from 18 -- 500 keV in eight time intervals. These spectra shown in Fig. 4 are all fitted with a simple power-law model for a comparison. In Table 2, we displayed the spectral parameters of fitting eight spectra with both a power-law model and a cutoff power-law model. From Table 2, we can clearly see the spectral variations of the source over time. In the simple power-law fitting results, we find the photon index varied from 0.7 -- 1.6, but the total flux from 20 -- 150 keV (around $8\times 10^{-11}$ erg cm$^{-2}$ s$^{-1}$) did not change significantly from 2003 -- 2011. If the cutoff power-law model was applied to the spectral fittings, the derived cutoff energies also changed with time, from $\sim$ 110 keV to higher than 250 keV.

\begin{table}

\caption{The hard X-ray spectral variations of the AXP 4U 0142+61 from 18 -- 500 keV in eight time intervals. Two different spectral models are applied to fit the spectrum: a single power-law model and a power-law plus the exponential cutoff model. The flux is given in units of $10^{-11}$ erg cm$^{-2}$ s$^{-1}$ in the range of 20 -- 150 keV.  }
\begin{center}
\begin{tabular}{l c c c c l}
\hline \hline
Time & Model &  $\Gamma$   &  Cutoff Energy (keV)  & Flux  &  reduced $\chi^2\ (d.o.f)$ \\
\hline
2003.12 &Power-law & $1.37\pm 0.21$ &  & $8.6\pm 1.1$ &  1.023(11) \\
 & Cutoff power-law &  $1.34\pm 0.62$ & $265.5\pm 120.9$ & $8.5\pm 1.2$ & 0.896(10) \\
\hline

2004.9 -- 2004.12 & Power-law & $1.00\pm 0.16$ &  & $7.3\pm 1.1$ &  1.498(11) \\
 & Cutoff power-law &  $0.70\pm 0.45$ & $161.2\pm 29.9$ & $7.4\pm 1.1$ & 1.116(10) \\
\hline
2005.6 -- 2005.7 & Power-law & $0.85\pm 0.12$ &  & $8.2\pm 1.0$ &  0.968(11) \\
 & Cutoff power-law &  $0.71\pm 0.31$ & $258.5\pm 100.4$ & $8.3\pm 1.0$ & 0.562(10) \\
\hline
2005.12 -- 2006.1 & Power-law & $1.07\pm 0.14$ &  & $8.7\pm 0.9$ &  0.794(11) \\
 & Cutoff power-law &  $0.92\pm 0.42$ & $260.4\pm 108.6$ & $8.8\pm 1.0$ & 0.365(10) \\
\hline
2006.7 -- 2006.7 & Power-law & $0.95\pm 0.17$ &  & $8.6\pm 0.8$ &  1.282(11) \\
 & Cutoff power-law &  $0.29\pm 0.32$ & $112.9\pm 39.7$ & $8.7\pm 0.9$ & 0.531(10) \\
\hline
2008.3 -- 2008.5 & Power-law & $1.29\pm 0.17$ &  & $8.1\pm 0.9$ &  0.986(11) \\
 & Cutoff power-law &  $1.01\pm 0.50$ & $191.14\pm 49.6$ & $8.2\pm 1.0$ & 0.835(10) \\
\hline
2010.1 -- 2010.10 & Power-law & $1.56\pm 0.23$ &  & $7.2\pm 1.0$ &  0.907(11) \\
 & Cutoff power-law &  $1.50\pm 0.61$ & $168.8\pm 47.9$ keV & $7.2\pm 1.1$ & 0.868(10) \\
\hline 
 2011.1 -- 2010.7 & Power-law & $1.00\pm 0.22$ &  & $6.9\pm 1.3$ &  1.201(11) \\
 & Cutoff power-law &  $0.99\pm 0.56$ & $195.1\pm 55.8$ keV & $6.9\pm 1.3$ & 1.230(10) \\
\hline

\end{tabular}
\end{center}
\end{table}

\section{Implications of Hard X-ray Cutoff}

The hard X-ray tails have been detected in several AXPs (e.g, 4U 0142+61, 1E 1841-0451, 1RXS J1708-40, see Kuiper et al. 2006,; Wang 2008), and also detected in magnetar/SGR activities (SGR 1900+14, SGR 1806-20, SGR 0501+4516 and 1E 1547.0-5408; see Esposito et al. 2007; Rea et al. 2009; Enoto et al. 2010b). During and after magnetar activities, the hard X-ray tails also show the hard spectra of photon index of $\sim 0.8 - 1.5$ (Rea et al. 2009; Enoto et al. 2010a, 2010b; Kuiper et al. 2012). Specially hard X-ray spectrum of the magnetar 1E 1547.0-5408 showed a hint of high energy cutoff around 100 keV during its 2009 activity (Enoto et al. 2010b). These hard X-ray properties during magnetar activities are still similar to that of the persistent emission in 4U 0142+61 we observed. The hard X-ray spectral property and cutoff feature of 4U 0142+61 can help us to understand the common properties and radiation mechanisms in different classes of magnetar candidates.

In fitting the average spectrum of AXP 4U 0142+61, three models are employed: power law,
bremsstrahlung, and cutoff power law.
4U 0142+61 is not detected by CGRO/COMPTEL observations (den Hartog et al. 2008).
Therefore, a power law without high energy cutoff is inconsistent with COMPTEL upper limits.
The bremsstrahlung model can naturally produce a cutoff in the hard X-ray spectrum ($E_{\rm cutoff}\approx kT$).
However, the bremsstrahlung model predicts a photon index $\Gamma \sim 1$ below the cutoff
energy for all sources. This is inconsistent with our current observations on available AXPs and SGRs
(different sources have different photon indexes, Gotz et al. 2006). Furthermore, the bremsstrahlung model
is of a hot spot form (e.g. the model of Thompson \& Beloborodov 2005; Beloborodov \& Thompson 2007).
However, observationally 4U 0142+61 shows hard
X-ray emissions in all phases (den Hartog et al. 2008). Therefore, the bremsstrahlung model is
not favored on physical ground. In our analysis of 4U 0142+61, the bremsstrahlung model
has a reduced $\chi^2$ of 4. Therefore, it is also not favored statistically.
Combining the nonthermal nature of magnetar hard X-ray emissions and the COMPTEL
upper limits, a cutoff power law is preferred for 4U 0142+61.

A cutoff power law for 4U 0142+61 results in a photon index $\Gamma \approx 0.5$ and cutoff
energy $E_{\rm cutoff} \approx 130 \,\rm keV$ (see Table 1). The photon index and cutoff energy
have relatively large uncertainties. Future hard X-ray observations may provide more accurate
measurements (e.g., by the Hard X-ray Modulation Telescope, HXMT). The present value of photon index
and cutoff energy can already give us some information about the hard X-ray emission mechanism of
magnetars. A cutoff energy around $130 \, \rm keV$ can rule out emission mechanism involving
ultra-relativistic electrons. The electron motion should be mildly relativistic (both microscopic and
bulk motions).

The quantum electrodynamics model of Heyl \& Hernquist (2005) predicts a cutoff energy much higher
than $1\,\rm MeV$. This is inconsistent with the cutoff energy of 4U 0142+61.
It also predicts a high energy gamma-ray flux detectable by Fermi-LAT. However, 4U 0142+61
is not detected by Fermi-LAT (Sasmaz Mus \& Gogus 2010; Tong et al. 2010). Therefore, the quantum
electrodynamics model can be ruled out with our current knowledge of magnetars.

The resonant inverse Compton scattering model by Baring \& Harding (2007) involves ultra-relativistic
electrons. It also predicts a cutoff energy much higher than $1\,\rm MeV$. Therefore, the ultra-relativistic
scattering model is inconsistent with our observations. Resonant inverse Compton scattering involving mildly relativistic
electrons is explored by Beloborodov (2013). It predicts a cutoff energy around $1\,\rm MeV$. While the cutoff
energy for 4U 0142+61 is around $130\,\rm keV$. Considering the uncertainty of cutoff energy due to the uncertainty
of photon index (a larger photon index will result in a higher cutoff energy), we can not rule out this model
at present. However, the problem with Beloborodov (2013) is that the particles there are from a transient corona.
The corresponding hard X-ray emissions will also be transient, with typical decaying time scale around one year.
The hard X-ray flux of 4U 0141+61 is stable over nine years as demonstrated in Section 4. Therefore,
if the hard X-ray emission of 4U 0142+61 is produced by mildly relativistic electrons, the electrons
must be from a persistent source.

AXPs and SGRs may be magnetars. At the same time they can also be neutron stars accreting from fallback disks
(Tong \& Xu 2011). The bulk motion of the accretion flow may also explain the hard X-ray emissions of 4U 0142+61
(Trumper et al. 2010). In the accretion case, the bulk motion of the accretion flow, the microscopic inverse Compton scattering,
and the resonant inverse Compton scattering\footnote{The magnetic field required by the resonant inverse Compton
scattering is only $10^{11}$--$10^{13} \,\rm G$ (Beloborodov 2013). Therefore, it can also happen in the accretion case.}
can all contribute to the hard X-ray emissions. A cutoff energy around $130 \,\rm keV$ requires that
both the bulk and microscopic motions of electrons should be at most mildly relativistic. Multiband observations of AXPs and SGRs
may help us distinguish between the magnetar model and the accretion model. X-ray polarization observations
may help us finally solve this problem.

\begin{figure}
\centering
  \includegraphics[width=0.5\textwidth,angle=270]{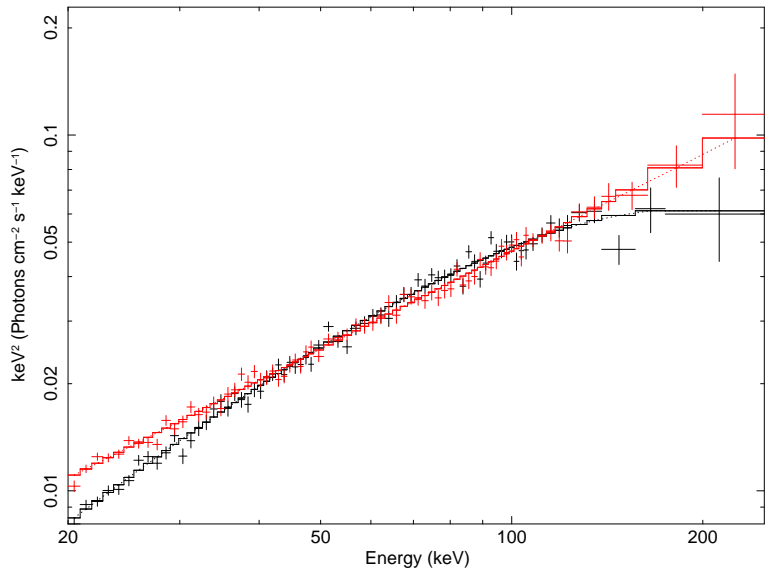}\\
  \caption{Simulated spectra of 4U 0142+61 for HXMT based on parameters of power law and cutoff power law models, shown in red and black respectively.  }\label{0142hxmt}
\end{figure}

Limited by sensitivity of INTEGRAL/IBIS, there is large uncertainty of cutoff energy due to
the uncertainty of photon index, which may be better constrained by future observations. Hard X-ray Modulation Telescope
(HXMT)\footnote{About the detailed information of this new mission, see the websites: http://www.hxmt.org/english, and http://heat.tsinghua.edu.cn/hxmtsci/book.html}, the first astronomical satellite to be launched by China, is a collimated hard X-ray telescope based on the direct demodulation method and NaI(Tl)/CsI(Na) phoswich detecting techniques. The payload consists of three telescopes; they work in the low, middle, and high energy range respectively, covering the 1--250 keV energy band. Among them the high energy telescope is sensitive between 20 and 250 keV, and has a large collecting area of 5000 cm$^2$ which provides high sensitivity. Based on parameters of
power law and cutoff power law model in Table 1, we simulate hard X-ray spectra of 4U 0142+61
with HXMT response matrix and background file, as shown in Fig. \ref{0142hxmt}. With
integration time of 1 Ms, there is obvious discrepancy between the two models below 30 keV
and around 200 keV, implying that future HXMT observations are able to provide more accurate
values for the photon index and cutoff energy, and verify the existence of cutoff in 4U 0142+61.


\section{Conclusion}

In this work, we systematically studied the hard X-ray spectral properties of a typical anomalous X-ray pulsar 4U 0142+61 based on a long-term hard X-ray monitoring of INTEGRAL. From 2003 -- 2011, the 20 -- 150 keV flux of 4U 0142+61 shows no significant variations. The derived hard X-ray luminosity of $\sim 10^{35}$ erg s$^{-1}$ above 20 keV is also similar to the luminosity in soft X-ray bands of 1 -- 10 keV. Though the total X-ray luminosity did not show variations, the hard X-ray spectral properties had some changes in the observed time intervals. If the spectra of 4U 0142+61 are fitted with a simple power-law model, the photon index $\Gamma$ changes from 0.7 -- 1.6. These spectra were also fitted by a cutoff power-law model, which gave $\Gamma\sim 0.3 - 1.5$ with cutoff energies $E_{\rm cutoff}\sim 110 - 250$ keV. To probe the high energy cutoff in the spectrum of 4U 0142+61, we accumulated all the data to obtain an average spectrum (with a higher significant level). A cutoff power-law model is preferred in spectral fittings, suggesting a spectral cutoff around 130 keV. It is the first time that the high energy cutoff of this AXP is derived, which will help us to understand the high energy radiation mechanisms of magnetars.

The detection of the high energy cutoff around 130 keV can exclude the resonant inverse Compton scattering model with ultra-relativistic
electrons (Baring \& Harding 2007). If the radiation still comes from magnetars, the mildly relativistic
electrons by persistent injections are required in the magnetar magnetosphere. However, the hard X-ray emission properties of 4U 0142+61 are still consistent with the accretion model, and the bulk and microscopic motions of electrons in the accreting flow should be mildly relativistic. Future hard X-ray observations specially multi-wavelength studies will further help to resolve the magnetar and accretion models. At least the present simulation studies suggest that future HXMT observations can provide better constraints on the high energy cutoff of 4U 0142+61 and more AXPs. In near future, good understanding of hard X-ray spectral properties of AXPs and SGRs by INTEGRAL and HXMT observations will provide important clues to nature and radiation mechanisms of magnetars.

\section*{Acknowledgments}

The authors would like to thank the referee for the comments and R. X. Xu for discussions. This work is based on observations of
INTEGRAL, an ESA project with instrument and science data center
funded by ESA member states.  W.W. is supported by the
National Natural Science Foundation of China (NSFC) No. 11073030. T.H. is supported by NSFC No. 11103021, West Light Foundation of CAS (LHXZ201201), Xinjiang Bairen project,
and Qing Cu Hui of CAS.

\end{document}